# THE IONIZATION PROFILE MONITORS IN THE RECYCLER RING *

B. Babacan[1, †], R. Ainsworth, K. J. Hazelwood, D. K Morris,
Fermi National Accelerator Laboratory, Batavia, IL
P. Snopok [1], [1]Illinois Institute of Technology, Chicago, IL

*Abstract*

The ionization profile monitors (IPMs) are used to measure the beam size in synchrotrons. Both the Fermilab Recycler and Main Injector (MI) machines have IPMs. However, they were not well understood enough to provide confidence in their measurements. Accurately measuring beam size through the IPMs was crucial to recognize the loss mechanisms for accelerators and to keep the beam loss to a minimum. Thus, performing measurements with different parameters using the IPMs led to a better analysis on how changes in conditions affect the beam size. The IPM measurements are compared with that of multi-wires in the upstream transfer line after applying corrections. The results were compared with other diagnostics and the change in the beam size for different parameters are presented in this paper.

## INTRODUCTION

The Recycler Ring (RR) is a key part of the Fermilab Accelerator complex used to deliver high intensity beams to the high energy neutrino experiments [1]. In order to understand the loss mechanisms that occur, it's crucial to have the ability to measure the transverse profile. Therefore, this paper mainly focuses on the ionization profile monitors (IPMs) [2] which can measure the beam size with the aim to have a better understanding of how this instrumentation operates and to have more confidence in its measurements.

The IPMs do not cause a beam loss unlike the multi-wires (MWs), and this makes them a better alternative to study the beam sizes in synchrotrons. They do not have the same limitations as MWs where the IPMs capture multiple turns in the RR, and this is one of the reasons to study them further to have a better understanding of the measurements they provide.

There are horizontal and vertical IPMs located in the RR and the Main Injector (MI) where the beams reach energies of 8GeV and 120GeV, respectively. For the purpose of this study, the IPMs are used to measure the beam size and the emittance of the beam. Then, the results were compared to those of MWs that are located in the RR and the MI-8 line which is the transfer line that connects the upstream machine, the Booster to the Recycler.

## IONIZATION PROFILE MONITORS

IPMs are used to measure the beam size and the position of the beam by looking at beam profiles of up to 64,000 individual turns of beam in a cycle. They use micro-channel plates (MCP) to receive data and to collect the excited particles through the electric and magnetic fields. A MCP is made from a resistive material, usually glass, up to 2.0 mm thick with an array of small tubes, called microchannels, passing from one face of the plate to the other face. The microchannels are typically 5 to 20 microns in diameter, parallel to each other and aligned at an angle of 8 to 13 degrees to the surface of the plate. These IPMs use two stacked MCPs rotated 180 degrees forming a "chevron" pattern to increase the gain and reduce the feedback in the MCP. The MCPs are controlled by voltage and the higher voltages can lead to degrading the quality of the MCPs over time.

This also leads to having unusable data points when individual pre-amplifiers for a channel fail giving disconnected data points within a profile. During data analysis, the channels that were marked to be inoperative were set to the average value of the overall IPM data set to eliminate the poor MCP issue.

Out of all 64,000 turns, both horizontal and vertical IPMs store the data locally but only return the first 1000 turns for analysis. This allows to calculate the sigma $\sigma$ that represents the beam size.

The IPMs were used to study the change in beam size in the MI by changing the MCP voltage to determine its effects as both were functioning compared to the vertical IPM not working in the RR. The R-square of the fits were also calculated to analyze the quality of the fits and this determined which voltage range was the best fit. Once an ideal MCP voltage range was determined, the beam size and the emittance were analyzed by using intensity as the dependent.

Afterwards, the emittance of the beam was calculated by using $\sigma$ in Eq. (1) where $\beta$ is a Twiss parameter, $D$ is dispersion, and $\frac{\delta p}{p_0}$ is the momentum spread.

$$\frac{\sigma_{x,y}^2}{\beta_{x,y}} - \frac{D_{x,y}^2}{\beta_{x,y}}\left(\frac{\delta p}{p_0}\right)^2 = \varepsilon_{x,y} \qquad (1)$$

The emittance was normalized by using the Lorentz factors $\beta_L$ and $\gamma$ in Eq. (2).

$$\varepsilon_{N,95\%} = 6\beta\gamma\varepsilon_{x,y} \qquad (2)$$

## MULTI-WIRES

The MW103 and MW104 are located in the Fermilab RR and MWs that start with an '8' are in the MI-8 line. Figure 1 displays the location of those MWs where the horizontal and vertical IPMs are next to MW104 and MW103, respectively.

---

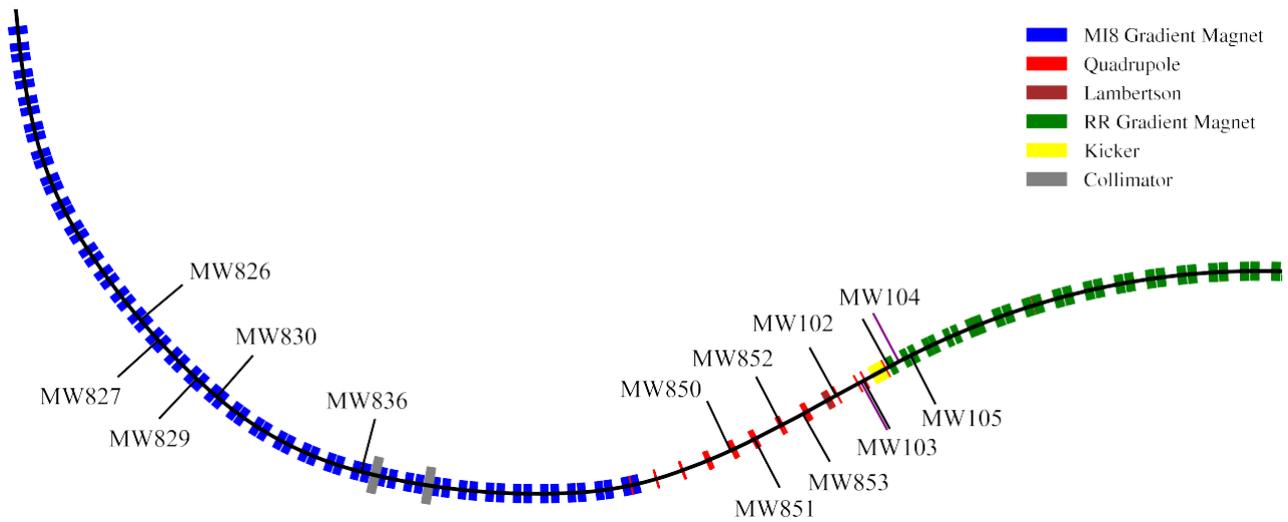

Figure 1: A schematic of the MI-8 line connecting to the Recycler ring. The horizontal IPM is next to MW104 and the vertical IPM is next to MW103.

Although the measurements of the MWs are precise, they also create a great beam loss. The MWs only record a single turn while IPMs can look at multiple turns without hurting the instrument. Therefore, MWs in the RR are only operated for one turn to avoid saturating the scanners and to prevent the beam from damaging the wire planes.

The horizontal and vertical MWs in MI-8 line measure the number of counts of a proton beam via the wire planes. A Gaussian fit was also applied to the number of counts collected on the wire planes and calculated the beam size using the same method as the IPMs. An example of MWs in the RR profile is shown in Fig. 2. The emittance was also calculated using Eq. (2) by using the corresponding Twiss parameters for each MWs.

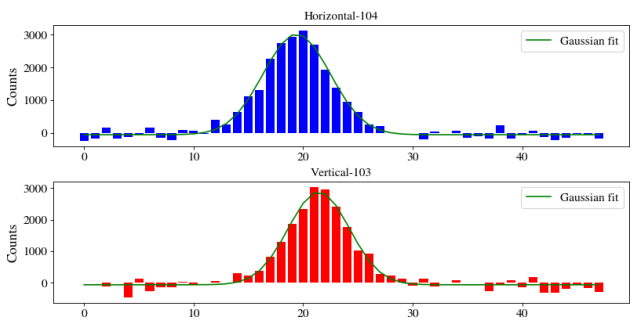

Figure 2: MW103 and MW104 profiles in the RR.

## IPM - MCP VOLTAGE SCAN

Different MCP voltage levels were set as dependent using the IPMs in the MI to see if there is a change in the measured profile. This range varied between 1170V-1220V and three data sets were collected at each MCP voltage level in both cases. This study was done by using the IPMs in the MI since both the horizontal and vertical work. This was done in order to determine if the MCP voltage had a significant effect on the measurements and to decide which MCP voltage level was advantageous to use for the horizontal IPM in the RR to be compared to the MWs.

The horizontal IPM was used to measure the beam size showed high uncertainties and the beam size was observed to be much larger than the average. Therefore, the horizontal IPM was not shown in this paper.

The beam size was observed to increase in the vertical IPM in the MI as shown in Fig. 3 and the uncertainty was a lot higher at low voltage levels. The MCP voltage of 1170V-1190V provided a beam size that changed between 1.5mm-2mm. This voltage range where the beam size remains almost stable with low uncertainty was shown to be the ideal MCP voltage to use when operating IPMs in both RR and MI.

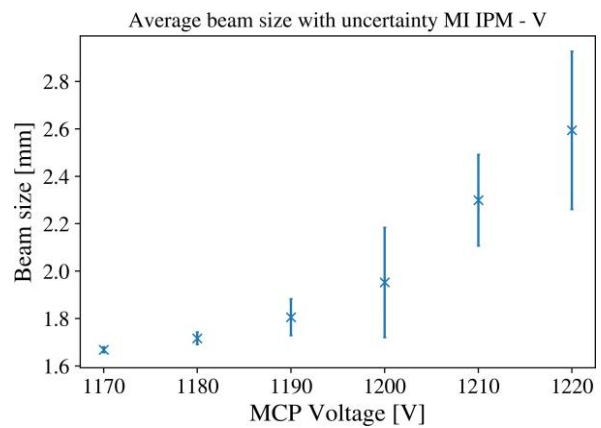

Figure 3: Beam size change in vertical MI-IPM.

Although, the MCP voltage should not be affecting the measurements in the IPMs, it was observed otherwise. It is also important to note that, the vertical IPM did not provide feasible beam size measurement at MCP voltage levels below 1170V.

## EMITTANCE MEASUREMENTS

Emittance is calculated for comparison because it represents the normalized beam size for any position in the lattice and not at a particular place. For both IPMs and MWs, their emittance were calculated using Eq. (2) where each have different $\beta$ and $D$ values as shown in Table 1. The term $\delta p/p_0$ was set to 5e-4 while calculating emittance.

Table 1: Twiss Parameters used for the MWs and IPMs

| Element | $\beta_x$ | $D_x$ | $\beta_y$ | $D_y$ |
| --- | --- | --- | --- | --- |
| 826 | 48.5 | -0.04 | 9.62 | -0.035 |
| 827 | 12.1 | 0.58 | 38.74 | -0.28 |
| 829 | 9.66 | 1.37 | 52.84 | -0.22 |
| 836 | 46.62 | 3.66 | 8.64 | -0.19 |
| 850 | 51.31 | -0.05 | 7.761 | 0.05 |
| 851 | 14.75 | 0.10 | 39.13 | 0.20 |
| 103 | 10.94 | 0.5 | 54.8 | 0.10 |
| 104 | 49.23 | -0.1 | 12.92 | 0.12 |
| IPMH | 27.10 | -0.32 | 23.308 | 0.19 |
| IPMV | 10.89 | 0.58 | 54.84 | 0.095 |

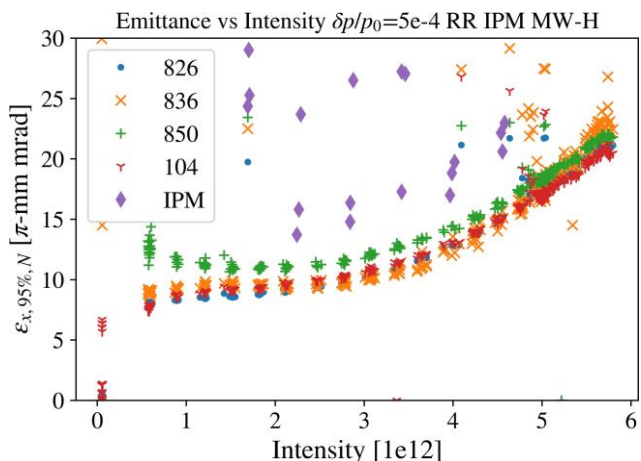

Figure 4: Emittance for Horizontal RR-IPM and MWs.

Some of the horizontal MWs were used to be compared to the IPM in the RR as listed in Fig. 4. This study involved an intensity scan for both IPMs and MWs and the emittance was shown to increase for both instruments. The horizontal IPM in the RR was operated at 1270V as it was in the range of the acceptable MCP voltage.

The emittance was calculated for the horizontal MWs as shown in Fig. 4. Due to the vertical IPM in the RR not functioning, no data was collected for the IPM and was not be able to compare it to the vertical MWs. The emittance for the vertical MWs was plotted in Fig. 5. The MW103 in the RR was shown to have lower emittance than the MWs in the MI-8 line.

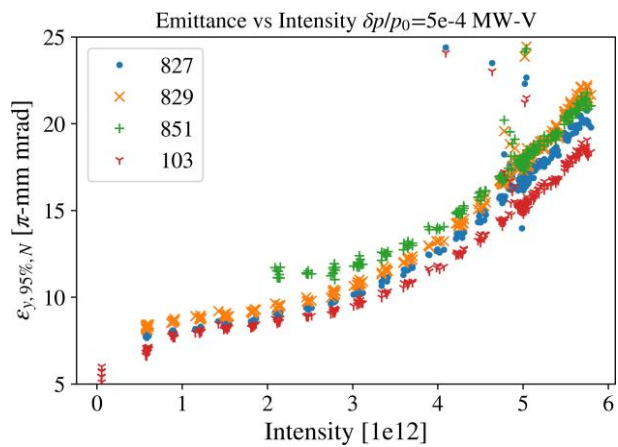

Figure 5: Emittance for vertical MWs.

## CONCLUSION

The beam size in the IPMs was observed to change with the MCP voltage and it sets a constraint on it depending on what beam intensity it is operated. At higher beam intensities, a lower MCP voltage level is required to provide a realistic beam profile and better Gaussian fits. It was also observed that operating IPMs at high MCP voltage and the beam at high intensities could saturate the IPMs enough to damage the MCPs. Therefore, it is important to use IPMs at certain MCP voltage levels to prevent damage and collecting poor quality data. This range was determined to be around 1170V-1190V.

The condition of the horizontal IPM in the MI should also be taken into consideration. It is older than the vertical IPM and that could affect the number of depleted MCPs. Hence, it could be the reason for high uncertainty. A comparison between IPMs in the RR and the MI would be beneficial when all IPMs are back to functioning simultaneously.

The horizontal MWs listed in Fig. 3 were shown to agree with each other while calculating the emittance. The emittance grows dependent on beam intensity in the RR and the MI. The emittance of the horizontal IPM in the RR was compared to those of MWs and shown that they do not quite agree with each other.

The emittance for the vertical MWs have shown agreement with each other and the horizontal MWs with similar values. The emittance became larger as the intensity increased.

A comparison of the vertical IPM to the MWs would be beneficial for future studies once it is functional again. It would be interesting to see how much the horizontal and vertical IPM compare with each other with the emittance calculations.

## ACKNOWLEDGEMENTS

Fermi Research Alliance, LLC manages and operates the Fermi National Accelerator Laboratory pursuant to Contract number DE-AC02-07CH11359 with the US DOE.